\documentclass[conference]{IEEEtran}

\def\BibTeX{{\rm B\kern-.05em{\sc i\kern-.025em b}\kern-.08em
    T\kern-.1667em\lower.7ex\hbox{E}\kern-.125emX}}

\usepackage{amsthm,amsmath,amsfonts}
\usepackage[inline]{enumitem}
\usepackage{caption,subcaption}
\usepackage{balance}
\usepackage[binary-units]{siunitx}
\DeclareSIUnit{\nothing}{\relax}
\usepackage{graphicx}
\usepackage{pifont}
\usepackage{array,booktabs}
\usepackage{multirow}
\usepackage{color}
\usepackage[normalem]{ulem} 

\usepackage{tikz,pgfplots,pgfplotstable}
\usetikzlibrary{decorations.pathreplacing}

\newtheorem{example}{Example}

\newcommand{\Replicas}{\mathcal{R}}
\newcommand{\Service}{\mathcal{S}}
\newcommand{\Clients}{\mathcal{C}}

\newcommand{\n}{\mathbf{n}}
\newcommand{\f}{\mathbf{f}}

\newcommand{\ID}[1]{\mathop{\textsf{DRF}}(#1)}
\newcommand{\Replica}[1]{R_{#1}}

\newcommand{\Client}[1][c]{\MakeLowercase{#1}}

\newcommand{\Name}[1]{{#1}}
\newcommand{\PName}[1]{{#1}}

\newcommand{\BFT}{BFT}
\newcommand{\CFT}{CFT}

\newcommand{\Paxos}{\PName{Paxos}}
\newcommand{\Cassandra}{\textsc{Cassandra}}

\newcommand{\Transaction}{\MakeUppercase{T}}
\newcommand{\MName}[1]{\textsc{#1}}
\newcommand{\Message}[2]{\textsc{#1}(#2)}

\newcommand{\Hash}[1]{\texttt{Hash}(#1)}

\newcommand{\LogDigest}[1]{\mathcal{L}_{#1}}
\newcommand{\Digest}[1]{\Delta_{#1}}
\newcommand{\Proposal}{\phi}

\newcommand{\ProposeTimer}{\tau_p}

\DeclareSIUnit\k{k}
\DeclareSIUnit\ms{ms}
\DeclareSIUnit\GB{GiB}
\DeclareSIUnit\B{B}
\DeclareSIUnit{\txn}{txn}
\DeclareSIUnit{\batch}{batch}
\newcommand{\RN}[1]{%
  \textup{\expandafter{\romannumeral#1}}%
}

\usepackage[noend]{algorithmic}
\newenvironment{myprotocol}{
    \hrule
    \smallskip
    \scriptsize
    \algsetup{linenosize=\tiny}
    \begin{algorithmic}[1]
        \newcommand{\SPACE}{\item[]}	
	
        \newcommand{\TITLE}[2]{\item[] \textbf{\underline{##1}} (##2) \textbf{:}\\[0.5pt]}
        \makeatletter
            \newcommand{\EVENT}[1]{\STATE \textbf{event} ##1 \textbf{do}\begin{ALC@g}}
            \newcommand{\ENDEVENT}{\end{ALC@g}}
        \makeatother
	
	\makeatletter
            \newcommand{\FUNC}[2]{\STATE \textbf{function} \textbf{##1} (##2) \begin{ALC@g}}
            \newcommand{\ENDFUNC}{\end{ALC@g}}
        \makeatother

	\newcommand{\INITIAL}[2]{\item[] \textbf{\underline{##1}} ##2\\[0.5pt]}
	
}{
    \end{algorithmic}
    \smallskip
    \hrule
}

\begin{document}

\title{Did we miss P In CAP? \\Partial Progress Conjecture under Asynchrony}

\author{\IEEEauthorblockN{Junchao Chen$^*$, Suyash Gupta$^\dagger$, Daniel P. Hughes$^\ddagger$, Mohammad Sadoghi$^*$}
\IEEEauthorblockA{
	\textit{$^{*}$Exploratory Lab, University of California, Davis} \\
	\textit{$^\dagger$University of Oregon} \\
	\textit{$^\ddagger$Radix DLT.}
}}

%
%

\maketitle

\begin{abstract}
Each application developer desires to provide its users with consistent results and an always-available system despite failures. Boldly, the CALM theorem disagrees. It states that it is hard to design a system that is both consistent and available under network partitions; select at most two out of these three properties. One possible solution is to design coordination-free monotonic applications. However, a majority of real-world applications require coordination. We resolve this dilemma by conjecturing that partial progress is possible under network partitions. This partial progress ensures the system appears responsive to a subset of clients and achieves non-zero throughput during failures. To this extent, we present the design of our \Cassandra{} consensus protocol that allows partitioned replicas to order client requests.

\end{abstract}

\section{Introduction}
More than two decades ago, when Brewer proposed the CAP theorem, it became evident to database 
practitioners and researchers that it is hard to design a system that is both consistent and available under network partitions~\cite{cap-theorem,cap-critique,cap-theorem-perspectives}.
Unsurprisingly, each application designer still wants to offer its client consistent results and an always-available system despite any failures~\cite{developer-concurrency,scalable-oltp-cloud}.

Node (or server) failure is the most common type of failure. 
To handle node failures ({\em availability}), application developers employ replication by deploying multiple replicas.
Moreover, an application's clients are often spread across the globe, which results in these replicas being distributed across the globe for low latency to the clients.
Having multiple replicas necessitates keeping these replicas in sync ({\em consistency}). 
To do so, these applications run consensus protocols like Paxos~\cite{paxos} and PBFT~\cite{pbftj}, which aim to establish an agreement among the replicas 
on the order of executing client transactions despite replica failures.

\begin{figure}[t]
    \centering
    \includegraphics[width=0.2\textwidth]{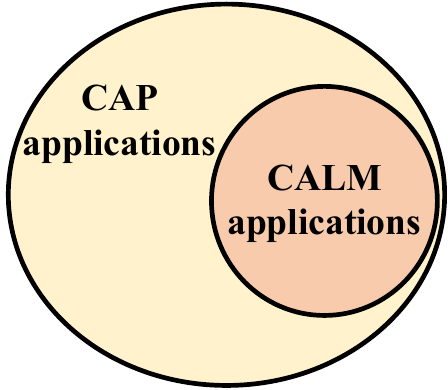}
    \caption{CAP vs. CALM applications.}
    \label{fig:calm-cap-venn}
\end{figure}

Unfortunately, replica failures are not the only common type of failure; the more catastrophic ones are network partitions, which 
hamper communication among the replicas, and thus consensus~\cite{flp}.
As a result, applications suffering from network partitioning are unable to make progress~\cite{network-reliable,network-partitioning-cloud}.
One way to ensure that an application is both consistent and available under network partitions is if it adheres to the CALM theorem,
which expects applications to be monotonic and lack any coordination~\cite{calm-theorem,calm-crdt}.
However, a majority of real-world applications require coordination and are often non-monotonic (refer to Figure~\ref{fig:calm-cap-venn}).

This makes us wonder, {\em if there exists a mechanism to allow non-CALM applications make progress under network partitions.}
In this paper, we show that under partitions, partial progress is possible for non-CALM applications.
Specifically, we conjecture that the ``P'' in CAP theorem could be broadened to capture partial progress instead of partition.
This partial progress guarantees that the application appears responsive to a subset of clients and achieves a non-zero throughput. 
Moreover, recovering from these failures requires replicas to run consensus on the pending requests, which can bottleneck the replicas 
and can lead to catastrophic failures~\cite{metastable-failures}.

To allow an application to make partial progress under network partitions, we present the design of a new fault-tolerant paradigm, \Cassandra{}.
Like existing \CFT{} protocols~\cite{paxos,raft} and \BFT{}~\cite{pbftj}, \Cassandra{} also requires a proposal selection phase and proposal commitment phase.
However, \Cassandra{} requires each replica to determine the strongest proposal among all the proposals and vote to support this strongest proposal.

Determining the strongest proposal is a two-step process:
(1) comparing rank of the proposals, and (2) comparing the suffix of the logs of committed transactions.
Each proposer runs a ranking function to assign its proposal a rank. 
Each replica marks a proposal as the strongest if its proposer has the most committed transaction. 
If two proposer's have equal set of committed transactions, select the proposal with the highest rank.
This two-step process of \Cassandra{} allows replicas to continue processing client requests during network partitions 
and merge their states once the network is synchronous.
On merging the states, if any conflict arises, \Cassandra{} provides simple rules to reconcile these conflicts in 
such a manner that at least one subset of client requests are preserved.

This paper aims to not only present \Cassandra{} but to initiate a discussion that partial progress is possible under network partitions. 
We show that \Cassandra{} can be adopted by existing applications to achieve non-zero throughput under failures.
Although \Cassandra{} follows the design of traditional \MName{Paxos} protocol, the ideas presented in this paper can be easily 
extended to modern \CFT{} protocols~\cite{raft,dpaxos,txn-process-hw-cloud}.
Similarly, one can attempt to extend \Cassandra{} to Byzantine failures, as recovering from Byzantine failures is more expensive as shown by literature on Byzantine-fault tolerant systems~\cite{chemistry,poe, geobft, ext_byshard, rcc, kang2024spotless, amiri2024bedrock}.

To illustrate our partial progress conjecture in practice, 
we deploy our \Cassandra{} protocol in the wild over the Twitter data.
We aim to capture the impacts of delays introduced by network partitions 
on social interactions such as tweeting (with mentions) and re-tweeting along with quoting, replying, and liking a tweet.
Our results show that once the network is restored, the system throughput peeks at $6\times$ 
of the steady state due to partial progress made during the partition.

\section{\Cassandra{} at a Glance}
From a bird's-eye view, \Cassandra{} looks like \Paxos{}. 
Like \Paxos{}, \Cassandra{} provides following desirable properties:
\begin{itemize}[nosep,wide]
\item {\bf Equal Leadership Opportunity.} 
There is no pre-designated leader; each replica can send a proposal.

\item {\bf Two-phase Linear Consensus.}
Each replica runs a two-phase protocol to accept a proposal.
\end{itemize}

Additionally, \Cassandra{} provides following new properties:
\begin{itemize}[nosep,wide]
\item {\bf Partial Progress under Network Partition.}
If replicas of a system are unable to communicate with each due to a partitioned or asynchronous network, 
\Cassandra{} allows these replicas to speculatively order and execute client transactions;
\Cassandra{} guarantees that transactions from at least one partition will persist.

\item {\bf Multi-Proposal Acceptance.}
\Cassandra{} supports partial-ordering, which allows it to accept non-conflicting proposals from multiple replicas.
\end{itemize}

\subsection{Preliminaries}
Prior to explaining the design of \Cassandra{}, we lay down the system model. 
We make standard assumptions also made by existing \CFT{} protocols~\cite{paxos,raft}.

We assume a replicated system $\Service = \{\Replicas, \Clients\}$, where $\Replicas$ 
denotes the set of replicas and $\Clients$ denotes the set of clients.
This replicated system has a total of $\n$ replicas, of which at most $\f$ are faulty;
$\n = 2\f + 1$.
The $\f{}$ faulty replicas can fail stop or crash; we do not assume any Byzantine failures.
We adopt the same partial synchrony model as existing \CFT{} protocols; 
safety is guaranteed under asynchrony, while liveness is only guaranteed during 
periods of synchrony.

\begin{figure}[t]
    \centering
    \begin{tikzpicture}[yscale=0.3,xscale=1]
        \draw[thick,draw=black!75] (0.75, 11.5) edge[green!50!black!90] ++(7.5, 0)
                                   (0.75,   8) edge ++(7.5, 0)
                                   (0.75,   9) edge[blue!80!black!60] ++(7.5, 0)
                                   (0.75,   10) edge ++(7.5, 0);

        \draw[thin,draw=black!75] (1, 8) edge ++(0, 3.5)
                                  (2, 8) edge ++(0, 3.5)
				  (3, 8) edge ++(0, 3.5)
                                  (4, 8) edge ++(0, 3.5)
                                  (5, 8) edge ++(0, 3.5)
				  (6, 8) edge ++(0, 3.5)
				  (7, 8) edge ++(0, 3.5);

        \node[left] at (0.8, 8) {$\Replica{3}$};
        \node[left,blue!80!black!60] at (0.8, 9) {$\Replica{2}$};
        \node[left] at (0.8, 10) {$\Replica{1}$};
        \node[left] at (0.8, 11.5) {$\Client{}$};

        \path[->] (1, 11.5) 	edge node[above,pos=0.8] {$\Transaction$} (2, 10)
				edge (2, 9)	
				edge (2, 8)
                  (2, 10) 	edge (3, 9)
                        	edge (3, 8)
		  (2, 9) 	edge (3, 10)
                        	edge (3, 8)
		  (2, 8) 	edge (3, 10)
                          	edge (3, 9)
                           
                  (3, 10) 	edge (4, 9)
		  (3, 9) 	edge (4, 9)
		  (3, 8) 	edge (4, 9)

		  (4, 9) 	edge (5, 10)
			 	edge (5, 9)
			 	edge (5, 8)

                  (5, 10)	edge (6, 9)
                  (5, 9) 	edge (6, 9)
                  (5, 8) 	edge (6, 9)

		  (6, 9) 	edge (7, 10)
			 	edge (7, 9)
			 	edge (7, 8)

		  (7, 10)	edge (8 ,11.5)
                  (7, 9) 	edge (8 ,11.5)
                  (7, 8) 	edge (8 ,11.5)
			;
        \node[below] at (2.5, 8) {\footnotesize \strut \Name{Propose}};
        \node[below] at (3.5, 8) {\footnotesize \strut\Name{Vote}};
	\node[below] at (4.5, 8) {\footnotesize \strut\Name{Prepare}};
	\node[below] at (5.5, 8) {\footnotesize \strut\Name{VotePrep}};
	\node[below] at (6.5, 8) {\footnotesize \strut\Name{Commit}};

    \end{tikzpicture}
    \caption{Schematic representation of the failure-free flow of \Cassandra{}.
We assume that $\Replica{2}$ has the strongest proposal.
	}
    \label{fig:cassandra}
\end{figure}
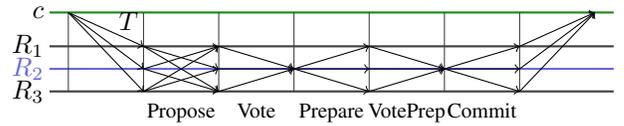

\subsection{Failure-Free Flow}
\label{s:civil}
A system is network partitioned if its replicas are divided into groups or partitions such that these partitions cannot exchange messages.
In such a setting, the system makes “partial progress” if the requests ordered by at least one partition commit once the network is synchronous.
Partial progress under asynchrony / network partitions illustrates that the system processes a subset of client requests and 
yields a {\em non-zero throughput}.
\Cassandra{} is the first consensus protocol to guarantee partial progress.

\Cassandra{} ensures partial progress by determining the {\em strongest proposer} for each round of consensus.
Finding the strongest proposer is a two-step process:
(1) comparing rank of the proposals, and
(2) comparing the suffix of the logs of committed transactions.
Each proposer locally runs a ranking function to assign its proposal a {\em rank}; higher the rank, greater are the chances of a proposal being selected.
The proposer sends the rank of its proposal along with the proposal.
Additionally, the proposer piggybacks its log of transactions.
A replica's log of transactions states the transactions it has voted, prepared, and committed, in order.
\Cassandra{} always selects the proposer with a larger log of committed transactions as the strongest proposer.

In Figure~\ref{fig:cassandra}, we schematically represent our \Cassandra{} protocol.
\Cassandra{} requires two phases to achieve consensus among its replicas.
The {\em first phase} aims to select the {\em strongest proposer (\S\ref{ss:strongest})}, 
while the {\em second phase} aims to commit the selected proposal.
\Cassandra{} makes use of {\em timers} in the first phase for rapid convergence (\S\ref{ss:convergence}); 
{\em \Cassandra{} is safe and live without timers}.
Each replica maintains an ordered log of transactions (denoted as $hist$).
Next, we describe \Cassandra{} in detail and present its pseudocode in Figure~\ref{alg:cassandra}.

{\bf Proposal.}
The first phase of \Cassandra{} starts when the $i$-th replica $\Replica{i}$ has a client transaction $\Transaction$ 
that it wants to propose (Line~\ref{alg:recv-client-txn}).
Next, $\Replica{i}$ creates a proposal $\Proposal_i$, which it broadcasts to all the replicas (Lines~\ref{alg:log-hash-create}-\ref{alg:broadcast-proposal}).
This proposal includes (1) hash of the ordered log $hist$, 
(2) client transaction and its hash, and 
(3) a function $\ID{}$ that assigns a rank to $\Replica{i}$'s proposal (\S\ref{ss:strongest}).
Post this, $\Replica{i}$ sets its proposal as the strongest proposal ($Max$) and waits on a timer ($\ProposeTimer$) to receive all the other proposals.

{\bf Proposal Selection.}
While waiting on the timer $\ProposeTimer$, replica $\Replica{i}$ may receive proposals (Line~\ref{alg:recv-proposal}) from several replicas.
It uses these proposals to determine the strongest proposer.
It compares the ordered log ($hist$) of each incoming proposal against $Max$ and sets $Max$ to the stronger proposal if any.
As described above, if two proposals have same ordered logs, then $\Replica{i}$ updates $Max$ to the proposal with higher rank (Line~\ref{alg:log-same}-\ref{alg:increase-max}).
Instead, if a received proposal's log has larger suffix, we update $Max$ to that proposal.

This implies that a replica with a lower rank but larger suffix gets higher preference than the replica with higher rank and smaller suffix.
We explain this choice of update in Section~\ref{ss:log-compare}.

{\bf Prepare.}
When replica $\Replica{i}$'s timer $\ProposeTimer$ timeouts, it sends a message $\MName{Vote}$ to the replica which has been selected as strongest proposer $Max$.  
Following this, each replica waits to receive the votes. 
Once a replica $\Replica{i}$ receives $\f{}+1$ matching $\MName{Vote}$ messages, it assumes that its proposal has been accepted by a majority of replicas.
Post this, $\Replica{i}$ asks all the other replicas to prepare themselves by broadcasting a $\MName{Prepare}$ message.

{\bf Commit.}
When a replica $\Replica{i}$ receives a $\MName{Prepare}$ message from the $j$-th replica $\Replica{j}$, 
it acknowledges this message by sending $\MName{VotePrep}$ message.
Once the replica $\Replica{j}$ receives identical $\MName{VotePrep}$ message from $\f{}+1$  replicas, it adds $\Transaction$ to its log $hist$ and 
broadcasts $\MName{Commit}$ message.
This $\MName{Commit}$ message allows all the replicas to eventually commit the transaction.

\begin{figure}[t]
    \begin{myprotocol}
	\INITIAL{Initialization:}{
	{\newline
	// $\ID{i}$ determines strength of $i$-th replica's proposal.\\
	// $hist$ is the ordered log of transactions.\\
	// $Max$ is the strongest proposal for a round.
	}}
	\SPACE

        \TITLE{Replica-role}{used by $i$-th replica $\Replica{i}$}
	\EVENT{Received a client transaction $\Transaction$ to propose} \label{alg:recv-client-txn}
		\STATE Compute hash of ordered log $\LogDigest{i} := \Hash{hist}$. \label{alg:log-hash-create}
		\STATE Compute hash of transaction $\Digest{i} := \Hash{\Transaction}$.
		\STATE Create proposal $\Proposal_i := \Message{Propose}{\ID{i}, \LogDigest{i}, \Transaction, \Digest{i}}$. 
		\STATE Broadcast $\Proposal_i$ to all the replicas. \label{alg:broadcast-proposal}
		\STATE Set $Max := \ID{i}$.
		\STATE Start a timer $\ProposeTimer$ \label{alg:being-propose-timer}
	\ENDEVENT
	\SPACE

	\EVENT{Received a proposal $\Proposal_j$ from $j$-th replica $\Replica{j}$} \label{alg:recv-proposal}
		\STATE Compare the suffixes of the two logs.
		\IF{$\LogDigest{j} \supset \LogDigest{i}$}
			\STATE $Max := \ID{i}$
		\ELSIF{$\LogDigest{j} = \LogDigest{i}$} \label{alg:log-same}
			\IF{$\ID{j} > Max$}
				\STATE $Max := \ID{i}$	\label{alg:increase-max}
			\ENDIF
		\ENDIF
	\ENDEVENT
	\SPACE

	\EVENT{Timer $\ProposeTimer$ timeouts}
		\STATE Send $\Message{Vote}{i, \Digest{Max}}$ to replica with $Max$ proposal.
	\ENDEVENT

	\EVENT{Received $\f{}+1$ matching $\Message{Vote}{i, \Digest{Max}}$ messages}
		\STATE Broadcast $\Message{Prepare}{i, \Digest{Max}}$
	\ENDEVENT
	\SPACE

	\EVENT{Received $\Message{Precommit}{i, \Digest{Max}}$ from $j$-th replica}
		\STATE Broadcast $\Message{VotePrep}{i, \Digest{j}}$
	\ENDEVENT

	\EVENT{Received $\f{}+1$ matching $\Message{VotePrep}{i, \Digest{j}}$ messages}
		\STATE Add $\Transaction$ to ordered log $hist$.
		\STATE Broadcast $\Message{Commit}{i, \Digest{j}}$
	\ENDEVENT

	\EVENT{Received $\Message{Commit}{i, \Digest{j}}$ from $j$-th replica}
		\STATE Add $\Transaction$ to ordered log $hist$.
	\ENDEVENT

    \end{myprotocol}
    \caption{\Cassandra{} protocol (failure-free path).}
    \label{alg:cassandra}
\end{figure}

\subsection{Deterministic Ranking Function}
\label{ss:strongest}
A key to success of \Cassandra{} lies in determining the strongest proposer among all the proposers.
Identifying the strongest proposal helps to converge the consensus faster as all the replicas would vote 
in the favor of such a proposal.
Doing so, however, faces the following challenge:
the strength function should be {\em fair} and should guarantee that {\em only one proposal} is marked as strongest in each round of consensus.

Existing literature presents several ways, which can be used to design such a strength function:
(1) Using a verifiable random function (VRF) to generate random numbers that have extremely low probability of collisions~\cite{vrf}.
(2) Proposal with highest priority or fees~\cite{pow-initial}, 
(3) Proposal that solves a puzzle fastest~\cite{blockchain-book, chen2022power}, and so on.
Although all of these designs yield desirable strength functions, they suffer from real-world adoption as 
no organization would like to pay for large compute and time.
As a result, in this paper, we design a simple, yet efficient {\em deterministic ranking function} (DRF).

We use DRF to assign each replica a unique rank in the range $\bf{[0,\n{}]}$.
Each replica calls the function $\ID{}$ to determine the rank of a replica.
The system administrator can decide the frequency of updating ranks of replicas, at the end of each consensus or 
post some interval of time.
In \Cassandra{}, higher the rank of a replica, higher is the chance of its proposal being accepted.

\begin{example}
Assume a system of $\n{} = 3$ replicas, where ranks of replicas $\Replica{1}$, $\Replica{2}$, and $\Replica{3}$ is 
$0, 2,$ and $1$, respectively.
Each replica broadcasts a new proposal during the proposal phase.
During the proposal selection phase, each replica receives proposals from all the replicas. 
As $\Replica{2}$ has the highest rank, each replica will select the proposal of replica $\Replica{2}$ as the $Max$ proposal, 
which will ensure that $\Replica{2}$ receives the necessary $\f{}+1$ votes.
\end{example}

\subsection{Rapid Convergence}
\label{ss:convergence}
A \CFT{} consensus protocol is only beneficial if it can guarantee eventual agreement of replicas on a single proposal.
In Lamport's \Paxos~\cite{paxos} protocol, it is hard to guarantee that replicas will ever converge, and thus, it cannot guarantee liveness.
Recent works~\cite{raft,dpaxos} eliminate this challenge by designating one replica as the {\em leader}.
As soon as a replica receives a proposal from the leader, it prepares itself for committing the leader's proposal.
Although designating one replica as the leader is the easiest solution, it suffers three challenges:
(1) It lacks fairness.
(2) It requires detecting and replacing the leader once it fails, which is not only expensive but hurts system throughput 
as no new transactions can be ordered until the new leader ensures that all the replicas have the same state.
(3) It prevents the system from making partial progress during a network partition if the leader is partitioned from the clients.

Due to these reasons, \Cassandra{} avoids designating any replica as the leader. 
Instead, it employs DRFs to assign each replica a rank, which helps in determining the strength of a proposal.
\Cassandra{} allows the rank of a replica to change over time; 
we can require each replica to run the DRF before each round of consensus.
Although DRFs guarantee fairness, they cannot eliminate the other two challenges.
We still need to provide uninterrupted transaction ordering under failures and network partitions. 
We argue that the use of timers is necessary for uninterrupted transaction processing.
We illustrate this through the following schemes, which do not require replicas to wait on a timer
(assume that Line~\ref{alg:being-propose-timer} in Figure~\ref{alg:cassandra} did not exist).
Allow of these schemes try to converge the replicas to a single proposal.
\begin{itemize}[nosep]
\item {\em Attempt 1.} 
We can ask all replicas to wait till they hear the proposal from the replica with the highest rank. 
Such a condition can cause a replica to hold off voting until it has seen all the proposals as the replica with strongest proposal may be slowest to broadcast 
or is suffering from message delays.
Worse, messages from the replica with the strongest proposal never arrive as it has crashed or partitioned, which will cause the system to get stuck.

\item {\em Attempt 2.}
We can ask each replica to vote for the first proposal that it receives. 
Such a solution has extremely low probability of success
because replicas are often spread across the globe to guard against data-center failures.
Unless at least $\f{}+1$ replicas receive the proposal from the replica with the strongest proposal prior to 
any other proposal, this solution will not lead to convergence.

\item {\em Attempt 3.}
We can ask a replica to vote for a proposal, once it has received proposals from a majority of replicas (in our case, $\f{}+1$).
Although this solution is better than Attempt 2, it still has a high probability of non-convergence as the first $\f{}+1$ proposals for 
a majority of replicas may not include the proposal from the replica with strongest proposal.
\end{itemize}

The non-convergence of these attempts makes us settle down for timers. 
Each replica initiates a timer after it receives a proposal, and once its timer expires, 
it compares the DRFs of these proposals to select the strongest proposal.
Although not all proposals may arrive until timeout, the system has a flexibility to {\em tune} the timeout value.
If replicas observe that the timeout period is large (all the proposals are arriving way earlier) and 
is increasing the latency of convergence, it can propose decreasing the timeout value. 
Instead, replicas can propose increasing the timeout value if they are unable to converge on a proposal.

\subsection{Determining Strongest Proposal}
\label{ss:log-compare}
One of the goals of \Cassandra{}, like most \CFT{} protocols, is to ensure that all the replicas agree to a common order for all the client transactions.
This common ordering choice impacts the way we can select a replica as the strongest proposer.
{\em Just simply because a replicas has strongest DRF does not guarantee that it has the strongest proposal}.
Consider the following example. 
\begin{example}
Assume a system of $\n{}=3$ replicas $\Replica{1}$, $\Replica{2}$, and $\Replica{3}$, such that all the replicas have committed and logged
transaction $\Transaction_0$.
Additionally, $\Replica{2}$ has committed $\Transaction_1$ and prepared $\Transaction_2$, while
$\Replica{3}$ has committed both $\Transaction_1$ and $\Transaction_2$.
Now, all the replicas are proposing for the subsequent round and assume the DRF sets the ranks for 
$\Replica{1}$, $\Replica{2}$, and $\Replica{3}$, as $3, 2, $ and $1$, respectively.
Despite having the strongest DRF, the proposal of $\Replica{1}$ never gets $\f{}+1$ votes as its log is missing entries for $\Transaction_2$  and $\Transaction_3$.
Although $\Replica{3}$ has committed $\Transaction_2$, $\Replica{2}$'s proposal gets $\f{}+1$ votes as it has both prepared $\Transaction_2$ and has a stronger DRF.
\end{example}

This example illustrates that selecting the strongest proposer requires analyzing the rank and suffix of the log of each proposer.
Given two replicas $\Replica{i}$ and $\Replica{j}$ with logs $hist_i$ and $hist_j$ and $\ID{i}$ and $\ID{j}$, respectively,
we generalize replica selection scheme (for setting $Max$ in Figure~\ref{alg:cassandra}) as follows:

\begin{enumerate}
\item If both the logs have same suffix, then $Max$ is set to the replica with the strongest DRF.

\item If suffix of $hist_i$ is $\{\Transaction_x \}$ and $hist_j$ is $\{\Transaction_x, Prep(\Transaction_y) \}$, such that $x < y$,
then $Max = \ID{j}$.

\item If suffix of $hist_i$ is $\{\Transaction_x \}$ and $hist_j$ is $\{\Transaction_x, \Transaction_y \}$, such that $x < y$,
then $Max = \ID{j}$

\item If suffix of $hist_i$ is $\{\Transaction_x, Prep(\Transaction_y) \}$ and $hist_j$ is $\{\Transaction_x, \Transaction_y \}$, such that $x < y$,
then $Max$ is set to the replica with the strongest DRF.
\end{enumerate}

In these rules, $Prep(\Transaction_y)$ implies that a replica has prepared transaction $\Transaction_y$, but has not received $\MName{Commit}$ 
message for $\Transaction_y$.

A keen reader would have observed that we require replicas to compare the suffixes of two logs when determining the stronger proposal.
Just comparing the suffixes is sufficient due to the following two reasons:
(1) Initially, each replica has the same state, and 
(2) The strongest proposer only marks its proposal as committed if it receives $\MName{VotePrep}$ messages from a majority ($\f{}+1$) of replicas. 
Once the strongest proposer marks its proposal as committed, there is a guarantee that in every set of $\f{}+1$ replicas in the system, 
there will be one replica that has committed the proposal.

\section{Network Partitions and Failures}
It is common for replicas to crash and get network partitioned~\cite{network-reliable}.
These failures not only impact a system's responsiveness to its clients but also have the potential of 
making the system stuck.
Before we illustrate how \Cassandra{} deals with failures, we try to classify all the failures 
that prevent consensus among the replicas.
\begin{itemize}[wide]
\item {\em Strongest Proposer Failure.}
The replica with the strongest proposal may fail at any time after it has been determined as the strongest proposer 
(receives $\f{}+1$ votes).
This could lead to replicas being in distinct states with respect to the strongest proposal, and 
\Cassandra{} needs to bring all these replicas to the common state.

For instance, consider the following worse-case scenario where the replica $\Replica{i}$ with the strongest proposal $\Proposal_i$ fails unexpectedly.
The timing of this failure can divide replicas into four non-empty sets.
Set $A$ includes less than $\f{}$ replicas that did not receive proposal $\Proposal_i$. 
Set $B$ includes more than $\f{}+1$ replicas that voted for $\Proposal_i$.
Set $C$ includes at least $\f{}+1$ replicas that have prepared $\Proposal_i$.
Set $D$ includes at most $\f{}$ replicas that have  committed $\Proposal_i$.

\item {\em Replica Failure.}
It is possible that a non-strongest proposer replica fails any time during consensus. 
This failing replica may be the only replica (apart from the strongest proposer) 
that has committed the strongest proposal. 
In such a case, \Cassandra{} needs to ensure that the committed proposal persists despite failures.

\item {\em Replica Partitioning.}
If the system starts experiencing network partitions, then its replicas may not be able to 
communicate with each other.
A network partitioning could lead to replicas being partitioned (divided) into two or more groups. 
Following this, the system can be in a state where either no group has at least $\f{}+1$ replicas, or
exactly one group has $\f{}+1$ replicas.
\end{itemize}

{\bf Failure Detection.}
\Cassandra{} aims to quickly detect failures, so that it can run recovery procedures.
To detect failures, in each round of consensus, we require each replica to maintain two additional timers: {\em vote timer} and {\em prepare timer}.
Each of these timers start at the end of the preceding phase and are used as follows:
(1) Each replica $\Replica{i}$ starts a vote timer once it sends a $\MName{Vote}$ message in support of the strongest proposal for that round.
When $\Replica{i}$ receives a $\MName{Prepare}$ message, it resets the vote timer.
(2) Each replica  $\Replica{i}$ starts a prepare timer once it sends a $\MName{VotePrep}$ message.
$\Replica{i}$ resets this prepare timer after it receives a $\MName{Commit}$ message.

If either of these timers expire, $\Replica{i}$ detects a failure and runs the {\em Merge} procedure, which we explain next.

\subsection{Merge Procedure}
\Cassandra's novel merge procedure aims to help replicas recover from the various types of failures described earlier.
It performs the following tasks:
(1) prevent committed state from being lost, and
(2) bring all the replicas to the common state.
As it is impossible to detect whether a replica has failed or it has partitioned (in an asynchronous network) our merge protocol 
provides a single algorithm to recover from these failures.
This requires us to slightly modify the rules that determine the strongest proposal (\S\ref{ss:log-compare}).

{\bf Modified Determining Strongest Proposal.}
These modified rules for determining a strongest proposal allow additional entries in the suffix of a log.
Specifically, a replica's log can include transactions that have neither prepared or committed. 
Later in this section, we explain why we allow adding such entries to the log.
For now, we denote these unprepared and uncommitted transactions as ellipsis ($\dots$) in the log.
Given two replicas $\Replica{i}$ and $\Replica{j}$ with logs $hist_i$ and $hist_j$ and $\ID{i}$ and $\ID{j}$, respectively,
the strongest proposer ($Max$ in Figure~\ref{alg:cassandra}) is set as follows:

\begin{enumerate}
\item If suffix of $hist_i$ is $\{\Transaction_x, ...\}$ and $hist_j$ is $\{\Transaction_x, ...\}$, then $Max$ is set to the replica with the strongest DRF.

\item If suffix of $hist_i$ is $\{\Transaction_x, ...\}$ and $hist_j$ is $\{\Transaction_x, ... , Prep(\Transaction_y), ...\}$, such that $x < y$,
then $Max = \ID{j}$.

\item If suffix of $hist_i$ is $\{\Transaction_x, ...\}$ and $hist_j$ is $\{\Transaction_x, ...., \Transaction_y, ...\}$, such that $x < y$,
then $Max = \ID{j}$

\item If suffix of $hist_i$ is $\{\Transaction_x, ...., Prep(\Transaction_y), ...\}$ and suffix of $hist_j$ is $\{\Transaction_x, ..., \Transaction_y, ...\}$, such that $x < y$,
then $Max$ is set to the replica with the strongest DRF.

\item If suffix of $hist_i$ is $\{\Transaction_x, ...., Prep(\Transaction_y), ...\}$ and suffix of $hist_j$ is $\{\Transaction_x, ..., Prep(\Transaction_z, ...\})$, such that $x < y \wedge x < z$,
then $Max$ is set to the replica with the strongest DRF.

\item If suffix of $hist_i$ is $\{\Transaction_x, ...., Prep(\Transaction_y), ...\}$ and suffix of $hist_j$ is $\{\Transaction_x, ..., \Transaction_z, ...\}$, such that $x < y \wedge x < z$,
then $Max = \ID{j}$.
\end{enumerate}

For instance, the first rule states the following: if the last transaction committed by two replicas is the same, 
then despite any set of unprepared transactions in the log of these replicas, the replica with the strongest DRF is selected as the strongest proposer.
A similar interpretation applies to the other rules.

The first four rules are straightforward extension of rules defined in Section~\ref{ss:log-compare}.
The last two rules are new and help in {\bf conflict resolution}.
The need for conflict resolution occurs when at least one replica has only prepared a transaction $\Transaction_y$, while another 
replica has instead prepared/committed a transaction $\Transaction_z$.
For instance, such a situation can arise if the network is getting repartitioned. 
In this situation, a replica $\Replica{i}$ only receives $\MName{Prepare}$ messages for $\Transaction_x$ (no $\MName{Commit}$ message) 
while another replica $\Replica{i}$ never prepares $\Transaction_x$ (as it did not receive $\MName{Prepare}$ messages for $\Transaction_x$) 
but commits $\Transaction_y$ as it receives both $\MName{Prepare}$ and $\MName{Commit}$ messages for $\Transaction_y$.

\subsubsection{Protocol Steps}
Next, we explain the Merge procedure that uses the aforementioned strongest proposal rules to allow system to make partial progress under failures.

{\bf Vote Timeout.}
If a replica $\Replica{i}$'s vote timer timeouts while waiting for the $\MName{Prepare}$ message, it assumes that the replica it voted 
as the strongest proposer has failed. 
As a result, it abandons this round (say $r$) of consensus and ignores any proposals from other replicas for round $r$. 
As $\Replica{i}$ was never selected as the strongest proposer in round $r$, 
it adds its own proposal for round $r$ to its log $hist$ and starts a new round of consensus ($r+1$). 
Notice that $\Replica{i}$ marks its proposal as {\em unprepared} in the log.
In round $r+1$, once $\Replica{i}$ has a new transaction to propose, it constructs a new proposal that extends its log $hist$ and follows 
the remaining steps in Figure~\ref{alg:cassandra}.

{\bf Prepare Timeout.}
If a replica $\Replica{i}$'s prepare timer timeouts while waiting for the $\MName{Commit}$ message, it assumes that the strongest proposer 
has failed after sending a $\MName{Prepare}$ message.
As a result, $\Replica{i}$ adds this strongest proposal to its log and marks its as prepared in its log. 
Next, $\Replica{i}$ terminates this round (say $r$) of consensus. 
When $\Replica{i}$ has a new transaction to propose, it constructs a new proposal that extends its log $hist$ and follows 
the remaining steps in Figure~\ref{alg:cassandra} to start a new round of consensus.

{\bf Conflict Resolution.}
During the proposal selection phase (\S\ref{s:civil}) a replica $\Replica{i}$ may receive a conflicting proposal $\Proposal_j$ from 
a replica $\Replica{j}$. 
This conflict is the result of $\Replica{i}$/$\Replica{j}$ preparing distinct transactions.
Fortunately, our rules for determining the strongest proposal dictate which replica's proposal should be considered strongest.
If $\Replica{i}$'s proposal is weaker, it rollbacks its state.

{\bf State Exchange.}
During the proposal selection phase (\S\ref{s:civil}) a replica $\Replica{i}$ may receive a proposal $\Proposal_j$ from 
a replica $\Replica{j}$ such that the suffix of log of $\Proposal_j$ ($hist_j$) is greater than the suffix of the log of current 
strongest proposal $Max$.
As per our aforementioned rules, $\Replica{i}$ should set $Max$ to $\Proposal_j$.
This implies that $\Replica{i}$ is ready to accept $\Replica{j}$'s log {\em as the most up-to-date history}.
As $\Replica{i}$'s log is missing some of the transactions, it queries $\Replica{j}$ for these missing entries.


\section{Evaluation}  
\label{s:eval}
To evaluate our partial progress conjecture in practice, we deploy our \Cassandra{} 
protocol in the wild over the Twitter data. We aim to capture social interactions such as tweeting 
(with mentions) and re-tweeting along with quoting, replying, and liking a tweet, all modeled with 
transactional semantics. More importantly, we further evaluate the impacts of delays introduced 
by network partitions that may occur due to possible adversarial attacks in order to disrupt the 
flow of information or by injecting misinformation (i.e., conflicting transactions) into a 
partitioned social network. Our aim is to study the effects on throughout and latency upon network 
recovery to demonstrate partial progress as the system heals itself.

\begin{figure}
\centering
\includegraphics[scale=0.2]{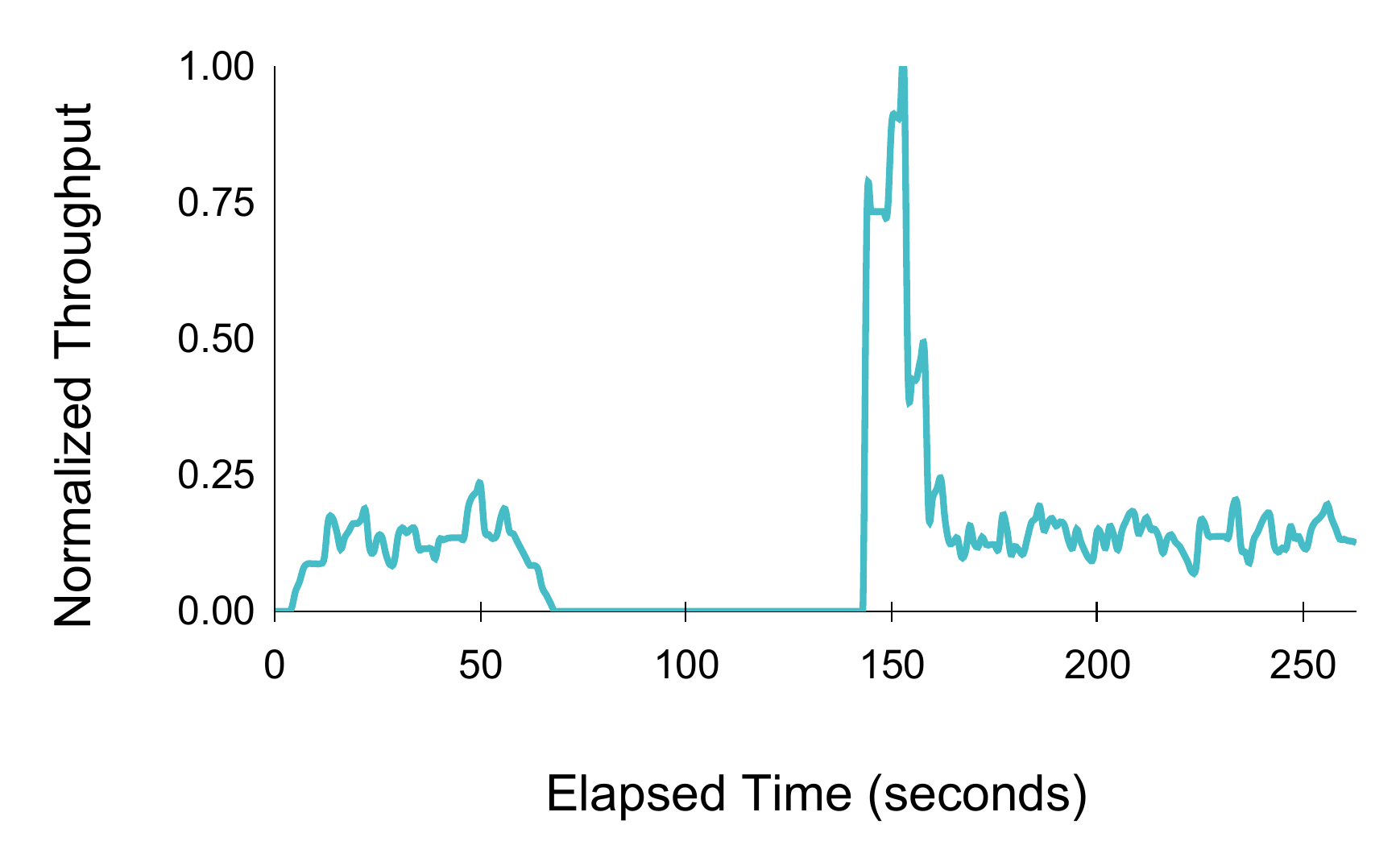}
\includegraphics[scale=0.2]{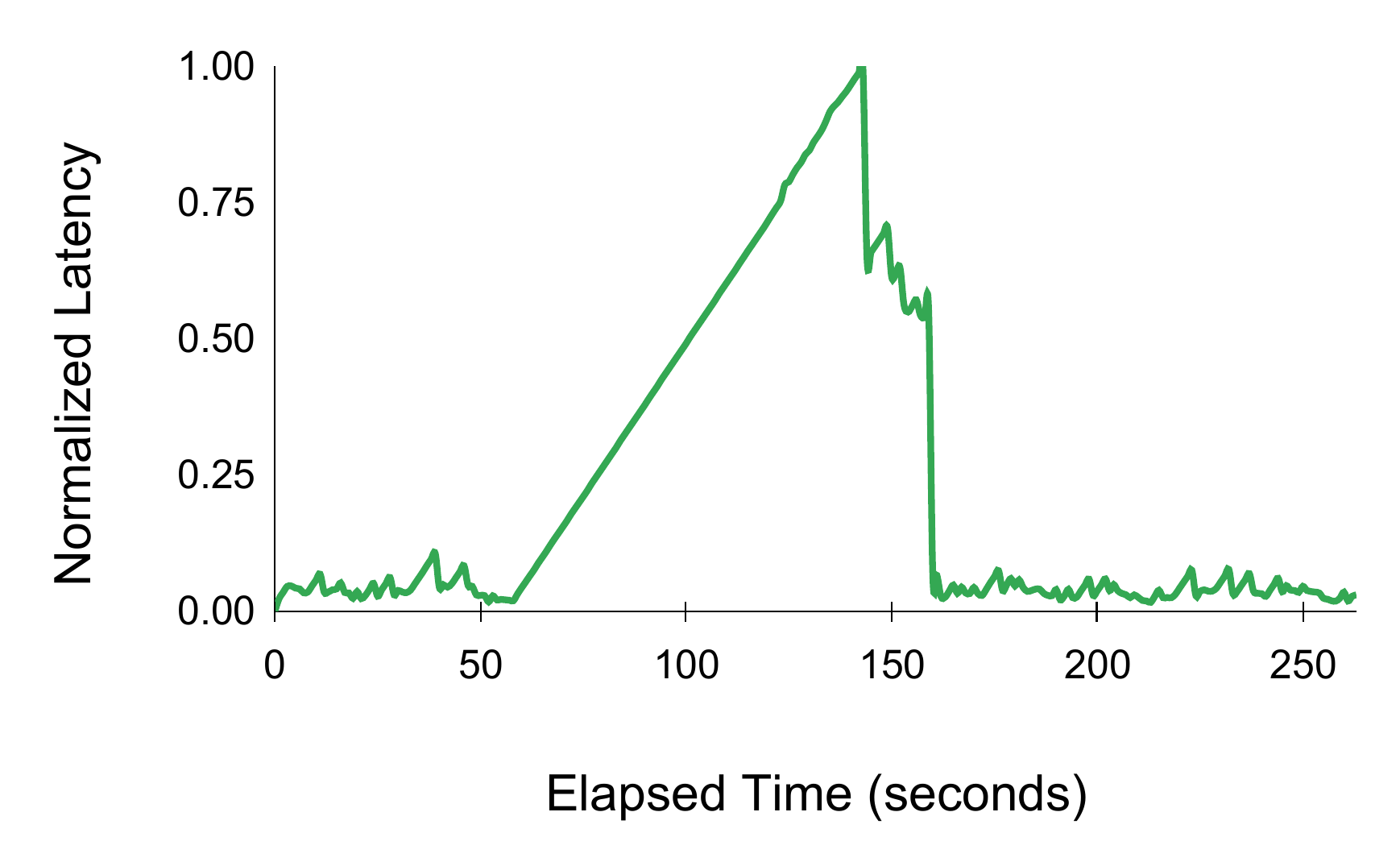}
\caption{Impacts of network partition on throughput and latency when partial progress is enabled.}
\label{fig:part}
\end{figure}

{\bf \em Setup and Benchmark.}
We run our experiments on a local cluster and deploy each replica on a virtual machine having a $4$-core CPU 
and $\SI{8}{\giga\byte}$ memory. We run \Cassandra{} on $100$ replicated machines, where the primary mode 
of communication is gossip to exhibit decentralized deployment. We inject 10,000 transactional tweets per 
second, where each transaction has a payload of under 1KB. The experiments are run for 260 seconds, where at 
the 60 seconds mark, the network undergoes partition, the partition lasts for 60 seconds, and the network is 
restored at 120 seconds. To collect results after reaching a steady state, we discard the measurement during the warmup 
period, and measurement results are collected over three runs.

{\bf \em Promise of Partial Progress.}
We evaluate the throughput and latency of the \Cassandra{} in partitioned setting where no partition has the 
majority of replicas. We observe during the partition, \Cassandra{} enters a speculative ordering and execution phase, 
where transactions are only softly committed and finality is only established once the partitions are recovered. 
Therefore, once the network is restored, we observe throughout peeks at $6\times$ of the steady state made possible by 
enabling partial progress. As expected, the finality latency is delayed proportionally to the 
length of the network outage. The results are shown in Figure~\ref{fig:part}.

\section{Related}
Brewer's CAP theorem~\cite{cap-theorem} argues that while designing an application, the application designer can only select two out of the following three properties: consistency, availability and partition tolerance.
Following this, several new models have appeared, which present fresh perspectives for an application designer.
PACELC~\cite{pacelc} captures the following double trade-off: if partitions occur, then trade between availability and consistency; else trade between latency and consistency.
FIT~\cite{fit} re-imagines the problem of a partitioned system as the problem of fairness and envisions fairness as a metric for latency.
CAC~\cite{tradeoffs-rep-system} redefines consistency as causal consistency and compares it against availability and convergence.
BASE~\cite{base} proposes a diametrically opposite model to ACID semantics. Unlike pessimistic nature of ACID transactions, BASE vouches for optimistic execution and accepts that the database consistency will be in a state of flux.
CALM~\cite{calm-theorem} proponates the ideas proposed by Conflict-replicated data types (CRDTs)~\cite{crdt}.
Prior works on designing fault-tolerant database systems include Google's Spanner~\cite{spanner} and Sequoia~\cite{sequoia}, which 
have been adopted in practice.
None of these works discuss the possibility of partial progress under CAP model, which is the theme of this paper, and we meet this aim with the help of our \Cassandra{} protocol.

\section{Conclusion}
In this paper, we illustrated that it is possible to design a system that is  consistent, available, and makes partial progress under network partitions.
We claim that this partial progress can help an application be responsive to a subset of its clients and allow it to achieve non-zero throughput.
To prove our claim, we present the design of our novel \Cassandra{} protocol that allows replicas to continue ordering 
client transactions even under failures.
\Cassandra{} achieves this goal by requiring each replica to vote for the strongest proposal among all the proposals.
Selecting the strongest proposal requires comparing the rank of a proposal and observing the suffix of its log.
Our experiments demonstrate that \Cassandra{} works in the wild and the partial progress during network partition helps to yield high throughputs once the network is restored.

\bibliographystyle{IEEEtran}
\bibliography{reference}

\end{document}